\documentstyle[12pt]{article}

\newcommand{\be}{\begin{equation}}
\newcommand{\ee}{\end{equation}}
\newcommand{\ba}{\begin{eqnarray}}
\newcommand{\ea}{\end{eqnarray}}

\begin{document}

\begin{titlepage}
\leftline{\_\hrulefill\kern-.5em\_}
\vskip -4 truemm
\leftline{\_\hrulefill\kern-.5em\_}
\centerline{\small{December 1993\hfill
Dipartimento di Fisica dell'Universit\`a di Pisa\hfill
IFUP-TH \ 62/93}}
\vskip -2.7 truemm
\leftline{\_\hrulefill\kern-.5em\_}
\vskip -4 truemm
\leftline{\_\hrulefill\kern-.5em\_}
\vskip1.5truecm
\centerline{Bosonization and the lattice Gross-Neveu model}
\vskip.7truecm
\centerline{
${\rm Matteo\:Beccaria}^{\,1,\,2,\,*}$}

\vskip1truecm
\centerline{\footnotesize{(1) Dipartimento di Fisica, Universit\'a di Pisa}}
\centerline{\footnotesize{Piazza Torricelli 2, I-56100 Pisa, Italy}}
\vskip1truemm
\centerline{\footnotesize{(2) I.N.F.N., sez. di Pisa}}
\centerline{\footnotesize{Via Livornese 582/a, I-56010 S. Piero a
Grado (Pisa) Italy}}
\vskip1truecm
\begin{abstract}
\small{
We consider a lattice version of the bosonized Gross-Neveu model. It
is explicitely chiral symmetric and its numerical simulation does not involve
any
anticommuting field.
We study its non trivial $1/N$ expansion up to
the next-to-leading term comparing the results with explicit numerical
simulations.
}
\end{abstract}
\vskip 1truecm
\hrule
\vskip 0.5truecm
$(*)$ beccaria@hpth4.difi.unipi.it, beccaria@vaxsns.sns.it
\end{titlepage}

\vskip .5 truecm
\noindent
{\bf 1.\,Introduction}

It is a well known property of two dimensional space-time that a free massless
fermion is equivalent to
a free massless boson~\cite{Coleman1}.
At the operator level, we can find a mapping, the Mandelstam
representation~\cite{Mandelstam},
which solves the fermionic field equations in terms of bosonic operators.
The vector and axial currents, together with
the energy momentum tensor satisfy the same Kac-Moody and Virasoro algebras in
the two
formulations~\cite{QCD2} allowing for a full correspondence.
The same renormalized Green functions are obtained for
the usual bilinear local fermionic operators and suitable local bosonic
operators.
The mapping can be extended to interacting models. Bosonization
is somewhat peculiar of $1+1$ dimensions where the spin assignment is arbitrary
but, on the other
hand, these models can help in studying physically interesting properties of
realistic theories
like confinement, chiral symmetry breaking and the baryonic spectrum.
{}From the point of view of the non perturbative analysis, we show in this
letter
how the bosonization technique is useful in
solving the problems encountered in the simulation of fermionic models.
The expensive inversion of the fermionic propagator is a serious obstacle for
numerical methods.
Moreover, for a large class of lattice models, the interplay between fermion
doubling and
chiral symmetry~\cite{Nielsen} complicates the
analysis of the continuum limit.
The bosonization approach is, in our opinion, powerful in both respects.
It avoids Grassmann fields {\it ab initio} and may
potentially overcome the Nielsen-Ninomiya theorem~\cite{Nielsen} as shown
in~\cite{Maggiore}.
There are also additional advantages concerning the so-called Symanzik
improvement, but
we postpone their discussion.
In this letter, we propose a lattice version of the two dimensional
Gross-Neveu model. It is a favoured theoretical laboratory for the analysis and
simulation
of fermionic models because it
has  nice analytical properties: it dynamically breaks a discrete chiral
symmetry
and is $1/N$ expandable allowing for an important check of the
simulation~\cite{Curci,Rossi1,Beccaria}. Its spectrum of bound states is
described
by a semiclassical formula from which mass ratios can be extracted.
We shall not consider here the problem of non-Abelian
bosonization~\cite{Witten2} and
the number of flavors $N$ will be only an expansion parameter. As a
consequence,
the largest part of the model's symmetry will be non linearly realized on the
bosonic fields.

\vskip .5 truecm
\noindent
{\bf 2.\,The continuum model}

The $1+1$ free massive fermion action may be bosonized
in terms of the Mandelstam representation~\cite{Mandelstam} which expresses
left and right handed Weyl fermions using a scalar massless field $\varphi(x)$:
\ba
\psi_L &=&  \sqrt{\frac{c\mu}{2\pi}}
: {\rm exp} \left\{-i\sqrt \pi\left(\int^x_{-\infty}d\xi\ \pi(\xi)+\varphi
(x)\right)\right\} : \ \ \ ,\\
\psi_R &=&  \sqrt{\frac{c\mu}{2\pi}}
: {\rm exp}\left\{-i\sqrt\pi\left(\int^x_{-\infty} d\xi\ \pi(\xi)-\varphi(x)
\right)\right\} : \ \ \ .
\ea
where $\pi(x) = \partial_0\varphi(x)$ and $c = \frac{1}{2}e^\gamma$
is related to the short distance expansion of the two dimensional
free bosonic propagator.
The normal ordering~\cite{Coleman2} is performed with respect to the mass
$\mu$~\footnote{We recall that in a continuum model with the cut-off scale
$\Lambda$, we have
\be
:\cos\beta\varphi : = \frac{\Lambda}{\mu} \cos\beta\varphi
\ee
where $\mu$ is an arbitrary renormalization scale.
}.
The bosonization dictionary is
\be
\label{dict}
\bar{\psi}{\partial \!\!\! /}\psi\to \frac{1}{2} (\partial_\mu\varphi)^2\qquad
\bar{\psi}\psi\to \frac{c\mu}{\pi} :\cos\beta\varphi : \qquad \beta^2 = 4\pi.
\ee
Actually, the above construction can be extended to
the massive Thirring model and Eqs.~\ref{dict}
correspond to the limit of vanishing interaction. In the general case, $\beta$
is a function of the
four-fermion coupling.
Our starting observation is that the purely fermionic Gross-Neveu model
may be formulated recasting the four fermion interaction in the form of
a space-time dependent mass term. This suggests a possible bosonization
procedure
which is particularly useful for an expansion in the number of flavours.
In the continuum, the fermionic Gross-Neveu model is described by the tree
level action
\begin{equation}
S = \int d^2x \left({\bar\psi}^{(\alpha)}
{\partial \!\!\! /}
\psi^{(\alpha)}-\frac{1}{2} g^2
\left({\bar\psi}^{(\alpha)}\psi^{(\alpha)}\right)^2 \right) \qquad \alpha =
1\cdots N,
\end{equation}
where $\psi^{(\alpha)}$ is a multiplet of two-dimensional spinors.
Rewriting the quartic interaction with a Lagrange multiplier (see~\cite{Rossi3}
for a
complete discussion of the model renormalization in this formulation) we obtain
\begin{equation}
S = \int d^2x \left({\bar\psi}^{(\alpha)}
{\partial \!\!\! /}
\psi^{(\alpha)}+\sigma
{\bar\psi}^{(\alpha)}\psi^{(\alpha)}+\frac{1}{2g^2}\sigma^2
\right).
\end{equation}
The above dictionary for bosonization can be used.
At the level of formal manipulations, the $\sigma$ field is just a
combinatorial device and our scheme is equivalent to Witten's treatment of the
bosonized Gross-Neveu model~\cite{WS,Witten1,Shankar}. We now propose a lattice
model
and show several indications
that it can be used to study the bosonic sector of the Gross-Neveu model, but
it must be kept
in mind that the precise nature and existence of its continuum limit is an open
problem.

\vskip .5 truecm
\noindent
{\bf 3.\,The lattice model}

We want to follow the analysis in~\cite{Curci}. Therefore, we write on the
lattice
\be
S = \sum_n\frac{N}{2\lambda}\sigma_n^2 + S_{\rm kin}(\Phi) + S_{\rm
int}(\sigma, \Phi)\ ,
\ee
where $\Phi$ is a set of fields which we now specify. For the fermionic model
of~\cite{Curci}, $\Phi$ is just a multiplet of $N$ Dirac fermions
$\psi^{(\alpha)}$,
$S_{\rm kin}(\psi)$ is some lattice kinetic action which avoids doubling and
\be
S_{\rm int}(\sigma, \psi) = \sum_{n, \alpha}\ \bar{\psi}^\alpha_n\
\psi_n^\alpha\ \sigma_n\ .
\ee
For the bosonized model, $\Phi$ is a set of $N$ massless bosonic fields
$\varphi^{(\alpha)}$,
$S_{\rm kin}(\varphi)$ is
\be
S_{\rm kin}(\varphi) = \frac{1}{2}\ \Delta_\mu^+\varphi_n^\alpha\
\Delta_\mu^+\varphi_n^\alpha
\qquad \Delta_\mu^+\varphi_n^\alpha =
\varphi_{n+\hat{\mu}}^\alpha-\varphi_n^\alpha
\ee
and
\be
S_{\rm int}(\sigma, \phi) = -\omega\sum_{n,\alpha}
\sigma_n\cos\beta\varphi_n^\alpha\ .
\ee
The constant $\omega$ comes from a tadpole computation. In a free scalar theory
with mass $\mu$
one can write
\be
\langle \cos\beta\varphi_n \rangle  = e^{-\frac{1}{2}\beta^2 \Delta_{n,n}} \ ,
\ee
and therefore
\be
\mu\ :\cos\beta\varphi:\ = \frac{1}{a} R(\mu) \cos\beta\varphi
\qquad
R(\mu) = \mu\exp\frac{\beta^2}{2}\int\frac{d^2 p}{(2\pi)^2}
\frac{1}{\hat{p}^2 + \mu^2}\ .
\ee
We take
\be
\omega = \frac{c\ R(0)}{\pi} = \frac{c\ 2^{5/2}}{\pi} \simeq 1.6035286\ .
\ee
The $Z_2$ symmetry
\be
\sigma\to-\sigma \qquad\qquad \varphi\to\varphi+\frac{\pi}{\beta}
\ee
is unbroken at the level of Feynman rules and
corresponds to the following symmetry of the fermionic realization
\be
\psi\to \gamma_5\psi\qquad\bar{\psi}\to -\bar{\psi}\gamma_5\qquad
\sigma\to -\sigma .
\ee
This symmetry is dynamically broken: the $\sigma$ field acquires a non zero
expectation value which
explains infrared stability.
As in the fermionic model, volume effects may be strongly reduced utilizing the
Symanzik improvement~\cite{Symanzik}.
Here comes another advantage of the bosonized model: the Symanzik improvement
adds no
more difficulties to the numerical simulation. On the other hand, in the
fermionic model,
the additional interactions reduce the extent to which the fermionic matrix is
sparse.
As a consequence, preconditioning becomes extremely more difficult.
As an aside, we remark that an alternative approach to the introduction of the
auxiliary field would be to bosonize
the purely fermionic model. The operator product expansion of the four-fermion
interaction
alters the kinetical term and fixes a trajectory in the $(\lambda, \beta)$
plane along which the
critical point is reached~\cite{Rossi2}.
In~\cite{WS,Witten1,Shankar}, this kind of bosonization has been applied
to the Gross-Neveu model in order to obtain important analytical results in the
continuum at
particular values of $N$. Here, we have chosen to introduce the field
$\sigma$ to mimic the $1/N$ expansion of the purely fermionic model.

\vskip .5 truecm
\noindent
{\bf 4.\,The $N=\infty$ limit}

At leading order, the effective action for the $\sigma$ field is
\be
\Gamma(\sigma) = \frac{\sigma^2}{2\lambda} - \frac{1}{L^2}\log\langle
e^{-S_{\rm int}}\rangle
\ee
where $L$ is the lattice side, $\sigma$ is kept space-time independent
and the vacuum expectation value is computed with respect to the free bosonic
action.
In the fermionic model, the interaction action is simply a mass term
and therefore the second term in $\Gamma$ can be explicitely evaluated
\be
\frac{1}{L^2}\log\langle e^{-S_{\rm int}}\rangle =
\frac{1}{L^2} \log\det(\bar{\Delta}_\mu\gamma_\mu + \sigma) =
\int\frac{d^2p}{(2\pi)^2} \log(\bar{\Delta}_\mu\gamma_\mu + \sigma) \ .
\ee
If we have instead
\be
S_{\rm int} = -\omega\sum_n \cos\beta\varphi_n
\ee
then a straightforward computation is not possible any more. However, we can
consider
\ba
{\rm F}_{\rm eff}(\sigma) &=& -\Gamma^\prime(\sigma) =
{\rm F}_{\rm cl}(\sigma) + {\rm F}_{\rm qu}(\sigma) \\
{\rm F}_{\rm cl}(\sigma)  &=& -\frac{\sigma}{\lambda} \\
{\rm F}_{\rm qu}(\sigma)  &=& \frac{\omega}{L^2}\langle \sum_n
\cos\beta\varphi_n\rangle
\ea
where the vacuum expectation value is taken now with
respect to the action of the $N=1$ model. A Monte Carlo calculation can give
very precise results without being too expensive.
We expect to find
\be
\lim_{\sigma\to\pm\infty} {\rm F}_{\rm qu}(\sigma) = \pm \omega \qquad
{\rm F}_{\rm qu}(\sigma) = -{\rm F}_{\rm qu}(-\sigma) \ .
\ee
On a finite lattice ${\rm F}_{\rm qu}(\sigma)$ has a finite slope as $\sigma\to
0$ which
signals the existence of a finite volume critical coupling for chiral symmetry
breaking.
Asymptotic scaling depends on the non trivial infinite volume, small $\sigma$
behaviour
of ${\rm F}_{\rm qu}(\sigma)$. We want to show that the solution of $F_{\rm
eff}(\sigma) = 0$
satisfies the scaling law
\be
\sigma = \Lambda_B \exp(-\frac{\pi}{\lambda})
\ee
This is true if we prove that the expansion
\be
{\rm F}_{\rm qu}(\sigma) = {\rm const}\ \sigma - \frac{1}{\pi} \sigma\log\sigma
+
O(\sigma^2\log^k\sigma)
\ee
does hold. However, a naive expansion of $\langle\cos\beta\varphi\rangle$
in powers of $\sigma$ fails when $L\to\infty$. The first terms are indeed
\ba
\langle\cos\beta\varphi\rangle &=& e^{-\frac{1}{2}\beta^2 J_0} +
\omega\sigma\sum_n
e^{-\beta^2 J_0}(\cosh\beta^2 J_n - 1) + \cdots \nonumber \\
&=& c_0(L) + \omega\sigma\  c_1(L) + \cdots
\ea
where
\be
J_n = \frac{1}{L^2}\sum_p e^{ipn}\frac{1}{\hat{p}^2}
\ee
computed introducing antiperiodic boundary conditions.
At a given $\sigma$ we cannot send $L\to\infty$ because $c_1(L)\to\infty$. The
question is: how can we relate the IR leading
logarithms of the $\sigma=0$, $L\to \infty$ limit to those of the
$\sigma\to 0$, $L=\infty$ one ?
Let us consider a function $F(\sigma, L)$ such that the $F(\sigma, \infty)$
exists and is
non analytic at $\sigma=0$. At fixed $L$, we can expand in powers of $\sigma$
\be
F(\sigma, L) = \sum_{\alpha\ge 0} c_\alpha(L) \sigma^\alpha
\ee
with vanishing radius of convergence as $L\to \infty$.
In this model, the volume effects can be kept constant uniformly in $\sigma$
by fixing the adimensional variable $S = \sigma L$. Therefore the leading IR
logs are
buried in the expansion
\be
G(\sigma) = \sum_{\alpha\ge 0} c_\alpha\left(\frac{k}{\sigma}\right)
\sigma^\alpha\ .
\ee
This simple approach works if the arbitrary constant $k$ is irrelevant for the
determination of
the logarithms. This is the case for $F_{\rm qu}(\sigma)$~\footnote{
Another example is the case of $c_0(L)$.
A numerical evaluation gives with great accuracy
\be
c_0(L) \sim \frac{\rm const}{L} \ .
\ee
We can assume
\be
\sum_p \frac{1}{\hat{p}^2} \sim {\rm const} + \frac{1}{2\pi}\log L \ .
\ee
Then, if one substitutes $L\sim\frac{1}{\mu}$, the correct logarithm in the
exact expansion of
\be
\int \frac{d^2p}{(2\pi)^2}\frac{1}{\hat{p}^2 + \mu^2}
\ee
as a function of $\mu$ is indeed obtained.
}.
A numerical investigation gives as $L\to\infty$
\ba
c_0(L) &\sim& \frac{\rm const}{L}\ ,\\
c_1(L) &\sim& \nu \log L\ .
\ea
With $\nu \simeq 0.12379219$ from which it follows that
$\omega^2 \nu$ is equivalent to $1/\pi$ up to our precision.
Besides this strong numerical indication of asymptotic scaling in the
$N=\infty$ limit, we give
a sketchy analytical confirmation that the slope is indeed the correct one. Let
us start from
\be
X = 2 e^{-\beta^2 J_0} \sum_n \sinh^2\frac{\beta^2}{2} J_n \ .
\ee
If we introduce a regularization mass $\mu^2$ and go to the continuum in the
second factor
\be
X \to 2 \exp-\beta^2\int\frac{d^2p}{(2\pi)^2}\frac{1}{\hat{p}^2 + \mu^2}
\cdot \int d^2 x \sinh^2\frac{\beta^2}{2}\langle\phi(x)\phi(0)\rangle \ .
\ee
The leading singularity in the $\mu\to 0$ limit is
\be
\langle\phi(x)\phi(0)\rangle \simeq -\frac{1}{2\pi}\log\sqrt{x^2}c\mu
\ee
whence, formally
\be
X = \frac{1}{2\pi^2\omega^2}\int d^2 x \frac{1}{x^2} \ .
\ee
We integrate this ill-defined quantity between the circles $\rho = a$ and $\rho
= L$
to pick up the volume dependence and get
\be
X = \frac{1}{\pi\omega^2}\log\frac{L}{a} \ .
\ee

\vskip .5 truecm
\noindent
{\bf 5.\,$1/N$ corrections}

The next to leading corrections in the $1/N$ expansion take the form
\ba
\Gamma(\sigma) &=& \frac{\sigma^2}{2\lambda} - \frac{1}{L^2}\log\langle
e^{-S_{\rm int}}\rangle + \\
& & + \frac{1}{2N}
\int \frac{d^2p}{(2\pi)^2} \log\left(\frac{1}{\lambda} + \Pi(p, \sigma)\right)
+
O(\frac{1}{N^2}) \ ;
\nonumber
\ea
$\Pi(p, \sigma)$ is the one particle irreducible
$\sigma$ self energy computed with the $N=1$ action in which
the dynamical field $\sigma$ is replaced by its $N=\infty$ value.
In the fermionic model, $\Pi(p, \sigma)$ is just a one loop lattice Feynman
diagram.
In the bosonized model, the numerical determination of $\Pi(p, \sigma)$ is not
straightforward.
We must utilize a Monte Carlo simulation to obtain the $\sigma$ field two point
function.
\be
-\omega^2(\langle\cos\beta\varphi(x)\cos\beta\varphi(y)\rangle -
\langle\cos\beta\varphi\rangle^2)
\ee
giving just $\Pi(x-y, \sigma)$.
By Fourier transforming, we obtain $\Pi(p, \sigma)$ and can compute the desired
$O(1/N)$
correction.
We then write
\be
\Gamma(\sigma) = \frac{\sigma^2}{2\lambda} +
\Gamma_1(\sigma) + \frac{1}{N} \Gamma_2(\sigma) + O(\frac{1}{N^2})
\ee
and expand in powers of $1/N$ the solution of $\Gamma^\prime(\sigma) = 0$:
\be
\sigma = \sigma_0 + \frac{1}{N} \sigma_1 + O(\frac{1}{N^2}) \ .
\ee
The desired correction is therefore
\be
\label{correction}
\sigma_1 = -\frac{\Gamma_2^\prime(\sigma_0)}
{1/\lambda - F_{\rm qu}^\prime(\sigma_0)} \ .
\ee
It is odd in $\sigma$ consistently with the unbroken discrete chiral symmetry.

\vskip .5 truecm
\noindent
{\bf 6.\,The simulation}

The simulation of the model has been performed using the Hybrid Monte Carlo
algorithm~\cite{hmc}.
As a first step, we have determined $F_{\rm qu}(\sigma)$ on a $40^2$ lattice
varying
$\sigma$ over a wide range of values using the $N=1$ model. The resulting plot
is shown in Fig. 1.
At $L=40$, the critical coupling for chiral symmetry breaking is about
$\lambda_c \simeq 0.38$.
Using the couplings $\lambda$
which can be extracted from that figure, we have studied the behaviour of
$F_{\rm qu}(\sigma)$ and $\Gamma_2(\sigma)$ around $\sigma = 0.2, 0.25, 0.3$ in
order to
determine with sufficient numerical accuracy the quantities needed to compute
$\sigma_1$.
Then, we have simulated the full model with $N=20$ flavours.
In Tab.~\ref{results}, we show at three values of $\lambda$, the leading $1/N$
prediction for
$\langle\sigma\rangle$, its $1/N$ correction, the $1/N$ corrected value
considering $N=20$ flavors
and the explicit numerical data including the measure of
$\langle\sigma^2\rangle_c = \langle\sigma^2\rangle - \langle\sigma\rangle^2$.
We see that the leading correction is necessary. The theoretical error on the
correction is
rather large; the discrepancy with the numerical data is therefore about $(8\pm
5)\%$.
All these simulations has been performed using the $40^2$ lattice and a
non-improved action.
We also simulated at the lower $\lambda = 0.49$ utilizing a $60^2$ lattice. We
did not
compute the theoretical $1/N$ corrections, but used this point just to look at
the behaviour of
\be
R(\lambda) = \langle\sigma\rangle
\exp\left(\frac{N}{N-1}\frac{\pi}{\lambda}\right)
\ee
which is also reported in Tab.~\ref{results}.
Having at our disposition two different lattice models,
we can relate the approximate finite renormalization shift in the
coupling to the $\Lambda$ parameters ratio. In~\cite{Curci}, we find the value
of $\lambda$ at
which the Symanzik action gives $\langle \sigma\rangle = 0.1488,\ 0.2,\ 0.25$
and $0.3$ at
$N=\infty$, $L=\infty$. Using
\be
\frac{1}{\lambda_B}-\frac{1}{\lambda_F} =
\frac{1}{\pi}\log\frac{\Lambda_B}{\Lambda_F}
\ee
and the result on the $N=\infty$, $L=40$ lattice for the bosonized model, we
obtain
$\Lambda_B = 90.6,\ 85.0,\ 80.7,\ 75.2$. The different results at different
$\lambda$ are due to
finite size effects at $L=40$ ($60$ for the first point) with the non-improved
action
and to non universal terms of the lattice $\beta$ function.
The values $|R - \Lambda_B|/\Lambda_B$, around $10\%$, may be explained with
the
$1/N$ corrections affecting $\Lambda_B$.

\vskip .5 truecm
\noindent
{\bf 7.\,Conclusions}

We have proposed a possible lattice action for a bosonization of the
Gross-Neveu model. There are many advantages with respect to the fermionic
formulation
from the point of view of the Monte Carlo simulation.
There are no anticommuting fields and fermion doubling can be naturally
avoided.
The computer codes are extremely fast and there is no need for any fine tuning
in order to
restore the $Z_2$ chiral symmetry.
In this letter we were not interested in studying scaling properties or the
mass spectrum of
the model.
Instead, we have shown the great simplicity and efficiency of its numerical
simulation including
its $1/N$ expansion.
A detailed description of the simulation including scaling results, a study of
the lightest
bosonic bound states and a thorough analysis of the Symanzik improved action
will be given elsewhere.
To be fair, we must remark once again that we limit our analysis to the bosonic
sector.
In our opinion, work on lattice actions for the bosonized ${\rm QCD}_2$ could
be extremely
interesting.

\vskip .5 truecm
\noindent
{\bf Acknowledgments}

I wish to thank Prof. Giuseppe Curci for constant encouragement. I am grateful
to
Prof. Paolo Rossi for many stimulating discussions and for a careful reading of
the
manuscript.

\eject

\begin{table}
\caption{Summary of the results}
\vskip 0.5truecm
\label{results}
\begin{tabular}{|c|cccccc|}
\hline
$\lambda$ & $\sigma_0$ & $\sigma_1$ & $\sigma_0 + \sigma_1/20$ & $\langle
\sigma\rangle$
& $\langle \sigma^2\rangle_c$ & $R$ \\
\hline
0.49000 & 0.1488 & $-$           & $-$            & $0.090(2)$ & $0.0606(3)$ &
$77(2)$\\
0.51905 & 0.2000 & $-1.7(2)$ & $0.12(1)$ & $0.127(1)$ & $0.0681(6)$ &
$74.6(6)$\\
0.54376 & 0.2500 & $-2.1(2)$ & $0.15(1)$ & $0.166(1)$ & $0.0744(6)$ &
$72.7(4)$\\
0.56876 & 0.3000 & $-2.4(2)$ & $0.18(1)$ & $0.209(1)$ & $0.0807(7)$ &
$70.0(3)$\\
\hline
\end{tabular}
\end{table}

\eject

\begin{figure}
\label{Fone}
\vskip 9truecm
\includegraphics{veff.ps}
\vskip.1in
\setbox0=\hbox{\noindent{ Figure 1:} $F_{\rm qu}(\sigma)$ on the $40^2$
lattice}
   	\setbox1=\vbox{\hsize=0.9\hsize\normalbaselines%
		\noindent{ Figure 1:} $F_{\rm qu}(\sigma)$ on the $40^2$ lattice}
   \ifdim\wd0<0.9\hsize%
	\medskip
        \centerline{\box0}%
   \else%
	\medskip
        \centerline{\box1}%
   \fi%
\end{figure}

\end{document}